\documentclass{pasj00}
 \tenpoint

\SetRunningHead{J.H. Fan et al.}{Polarization in  BLs}
\title{Radio Polarization of BL Lacertae objects}
\author{Jun-Hui Fan$^{1,2}${\thanks{email:fjh@gzhu.edu.cn}}, Tong-Xu Hua$^1$,  Yu-Hai Yuan$^{1}$, Yong-Xiang Wang$^{3}$,
 Yi Liu$^1$, \\
 Jiang-Bo Su$^{1}$, Yong-Wei Zhang $^1$, Jiang-He Yang$^4$, Yong Huang$^1$}
\affil{1. Center for Astrophysics, Guangzhou University, Guangzhou 510006, China \\
2. Physics Institute, Hunan Normal University, Changsha, China \\
3. College of Science and Trade, Guangzhou University, Guangzhou
511442, China\\
4.  Department of Physics and Electronics Science, Hunan University
of Arts and Science, Changde 415000, China}

\Received{$\langle$reception date$\rangle$}
\Accepted{$\langle$acception date$\rangle$}
\Published{$\langle$publication date$\rangle$}
\begin{document}

\KeyWords{BL Lacertae object, Polarization, Beaming Model}
\maketitle

\begin{abstract}

In this paper, using the database of the university of Michigan
Radio Astronomy Observatory (UMRAO) at three (4.8 GHz, 8 GHZ, and
14.5 GHz) radio frequencies, we studied the polarization properties
for 47 BL Lacertae objects(38 radio selected BL Lacertae objects, 7
X-ray selected BL Lacertae, and two inter-middle objects (Mkn 421
and Mkn 501), and found that
 (1) The polarizations at higher radio frequency is higher
 than those at lower frequency,
 (2) The variability of polarization at higher radio frequency is
 higher than those at lower frequency,
 (3)  The polarization is correlated with the radio spectral index,
 and
 (4) The polarization is correlated with
  core-dominance parameter for those objects with known core-dominance parameters
    suggesting that the relativistic beaming could explain the
polarization characteristic of BL Lacs.
\end{abstract}

\section{Introduction}

 BL Lacertae objects are generally described as a subclass of active
 galactic nuclei (AGNs), showing rapid and large amplitude variation,
 variable and high polarization, and core-dominated non-thermal continuum.
 Some BL Lacertae objects show superluminal motion and high energy
 gamma-ray emissions (
  Andruchow et al. 2005;
  Angel \& Stockman 1980;
  Fan 2005;
  Fan et al. 1996;
  Gabuzda 2003;
  Gu et al. 2006;
  Gupta et al. 2004;
  Qian et al. 2003, 2004;
  Romero et al. 1995, 2002;
  Stickel et al. 1993;
  Tao \& Qian, 2004;
  Wills et al. 1992;
  Xie et al. 2004).

  According to the surveys, BL Lacertae objects can be divided into
 radio selected BL Lacertae objects (RBLs) and the X-ray selected BL
 Lacertae objects (XBLs). From the spectral energy
 distribution (SED), one can see that some BL Lacertae
 objects show lower-frequency peaked SED and are called as
 LBLs, while some others show higher-frequency peaked SED and are
 called as HBLs (see Urry \& Padovani 1995 for a review). Generally,
 XBLs correspond to HBLs while RBLs to LBL. For Mkn 421 and Mkn 501,
 they are classified as RBLs by someone and as XBLs by some other
 authors. However, from SED, they are both classified as HBLs.

 The observational properties of RBLs are systematically different from
 those of XBLs.  The latter have flatter spectral energy distribution
 from the radio through X-ray, a higher observed peak of the emitted
 power from radio through X-ray spectral energy distribution. Furthermore,
 RBL and XBL show different observation properties(see
 Angel \& stockman 1980;
 Efimov, et al. 1988a,b, 2002;
 Fan et al. 1997;
 Giommi et al. 1995;
 Junnuzi et al. 1994;
  Sambruna et al. 1996).

 In this paper, we will investigate the radio properties of polarization in
BL Lacertae objects based on the MURAO data base. It is arranged as
follows.
 In section 2, we will calculate the averaged polarization and the variation of polarization for RBLs and XBLs.
  In section 3, we give some discussion and a conclusion.

\section{Calculation and Results}

From the MURAO data base, we can get 47 BL Lacertae objects, 38 of
them are RBLs, 7 of them are XBLs, two of them (Mkn 421 and Mkn 501)
are intermiddle type. For those sources, we did following
calculations

\subsection{Radio Spectral Index}

For each source, firstly we got its averaged flux densities at three
radio frequencies (4.8GHz, 8.0GHz and 14.5 GHz) respectively, then
adopted the linear regression analysis to the averaged flux at the
three frequencies,
   we  got the fitting result as $\log S_{\nu} = -\alpha \log \nu + c$.
  Therefore, we finally
  got the spectral
    index, $\alpha$
($S_{\nu}\propto\nu^{-\alpha}$). The
 resulting spectral indexes are
listed in
   Table 1, from which,
   we found that the radio spectral
indexes for the 7 XBLs are in the range of  $\alpha$ = -0.188 to
$\alpha$ = 0.625 with an averaged value of $\left<\alpha\right> =
0.235\pm0.255$ while those  for the 38 RBLs are in the range of
$\alpha$ = -0.601 to $\alpha$ = 0.757 with an averaged value of
$\left<\alpha\right> = 0.044\pm0.264$.

\subsection{Radio Polarization}

For the RBLs and XBLs, we calculated respectively the averaged
maximum polarization and the averaged polarization based on the
    observed
   polarization of each source.
     The detail procedure is as follows.
At each frequency for each source, we chose the observed maximum
polarization value as the maximum polarization at the corresponding
frequency, and calculated the averaged value of the whole
observation polarization as the averaged polarization. Therefore, we
have three maximum polarizations and three averaged polarizations
(at 4.8, 8.0 and 14.5 GHz) for each source. Then for each subclass
(XBLs and RBLs), we can calculate the averaged maximum polarization
at frequency $\nu$ ($\nu= 4.8$, 8.0 and 14.5 GHz),
 $P_{\nu}^{Max}=\Sigma P_{i, \nu}^{Max}/N$, here $P_{i,\nu}^{Max}$ is the maximum
observed polarization at frequency $\nu$ for the $i$th source in the
subclass, and $N$ is the source number of the subclass. $N = 7$ and
38 for XBLs and RBLs respectively. The averaged averaged
polarization is also calculated as $P_{\nu}^{Ave}=\Sigma
P_{i,\nu}^{Ave}/N$, here $P_{i, \nu}^{Ave}$ is the averaged observed
polarization for the $i$th source in the subclass.
 So, we
 can get following statistical results.

 $$\left<P_{4.80\rm{GHz}}^{\rm{Max}}(\%)\right> = 36 \pm 29,$$
 $$\left<P_{8.00\rm{GHz}}^{\rm{Max}}(\%)\right> = 65\pm 23,$$
 $$\left<P_{14.5\rm{GHz}}^{\rm{Max}}(\%)\right> = 74\pm 25,$$
and
 $$\left<P_{4.80\rm{GHz}}^{\rm{Ave}}(\%)\right> = 7\pm 5,$$
 $$\left<P_{8.00\rm{GHz}}^{\rm{Ave}}(\%)\right> = 16\pm 7,$$
 $$\left<P_{14.5\rm{GHz}}^{\rm{Ave}}(\%)\right> = 17\pm 7$$
for 7 XBLs,

 $$\left<P_{4.80\rm{GHz}}^{\rm{Max}}(\%)\right> = 17\pm 13,$$
 $$\left<P_{8.00\rm{GHz}}^{\rm{Max}}(\%)\right> = 26\pm 21,$$
 $$\left<P_{14.5\rm{GHz}}^{\rm{Max}}(\%)\right> = 29\pm 21,$$
 and
 $$\left<P_{4.80\rm{GHz}}^{\rm{Ave}}(\%)\right> = 4\pm 2,$$
 $$\left<P_{8.00\rm{GHz}}^{\rm{Ave}}(\%)\right> = 5\pm 3,$$
 $$\left<P_{14.5\rm{GHz}}^{\rm{Ave}}(\%)\right> = 6\pm 3$$
for 38 RBLs.

\subsection{Results}

   For polarization, the averaged
  results suggest that both the maximum
and averaged polarizations increase with the radio frequency in the
radio range of from 4.8GHz to 14.5GHz. In addition, within that
radio frequency range, both the averaged maximum and the averaged
averaged polarization in XBLs are  higher than those in RBLs at the
same radio frequency.

\begin{figure}
\begin{center}
    \FigureFile(90mm,100mm){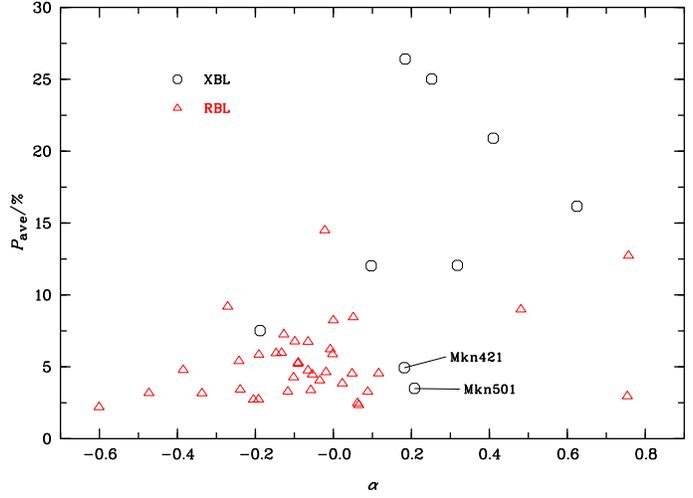}
\end{center}
\caption{ The relation between the
averaged polarization at 8 GHz and the radio spectral index for
radio selected BL Lacertae objects--RBL (triangles) and X-ray
selected BL Lacertae objects--XBL (open circles)}
\label{Fan-2006-PASJ-fg-Pa}
\end{figure}

   For the polarization and the spectral index, there
 is a
tendency for the radio polarization to increase with the spectral
index for BL Lacertae objects as shown in Fig.
\ref{Fan-2006-PASJ-fg-Pa}.

\begin{figure}
\begin{center}
    \FigureFile(90mm,100mm){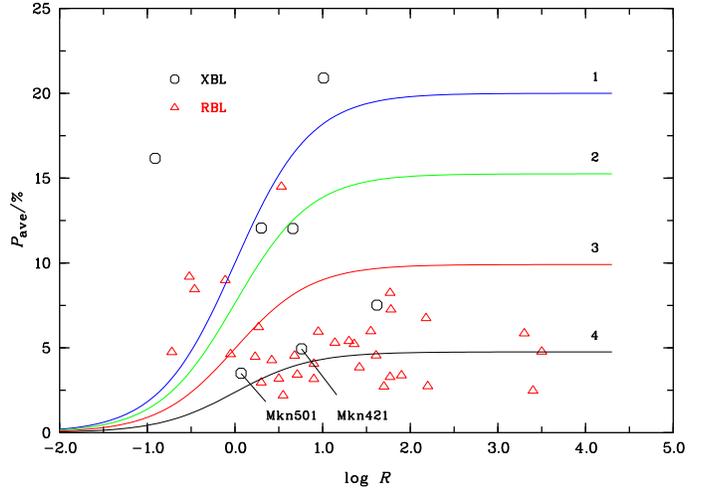}
\end{center}
\caption{ The relation between the
averaged polarization at 8 GHz and the core-dominance parameter for
radio selected BL Lacertae objects--RBL (triangles) and X-ray
selected BL Lacertae objects--XBL (open circles). Curve 1 stands for
$\eta$ = 0.25,
 curve 2 stands for $\eta$ = 0.18,
 curve 3 stands for $\eta$ = 0.11, and
 curve 4 stands for $\eta$ = 0.05.}
\label{Fan-2006-PASJ-fg-PR}
\end{figure}

  The core-dominance parameter is an important parameter
calculated from the radio observation. It is defined as the ratio of
the core flux density to the extended flux density (Orr \& Browne
1982).
 For polarization and the core-dominance parameter, we plotted
the averaged 8GHz polarization against the core dominance parameter
as shown in Fig. \ref{Fan-2006-PASJ-fg-PR}.

\section{Discussion}

It is believed that the particular properties observed from BL
Lacertae objects are associated with the beaming effect, and the
beaming model was adopted to explain both the particular
observational properties and some observational differences between
RBLs and XBLs (see
 Fan et al.  1997;
 Fan \& Xie 1996;
 Georganopoulos \& Marscher 1999)
   although the viewing angle alone can not explain all the
difference between
 RBLs and XBLs (Sambruna et al. 1996).

High and variable polarization is one of the characteristics of BL
Lacertae objects. For the radio polarization, although the averaged
value indicates that the polarization in
   XBLs
 is higher than that in
RBLs,
   we can
get the conclusion that the radio polarization in XBLs is not always
higher than that in RBLs.  The reasons are that 1) There are only 7
XBLs in our consideration, and 2) for the sources with the averaged
polarization being   higher than 10\%, we have two
   RBLs but
6 XBLs. However, for the  sources with maximum polarization being
   higher
than 35\%, we have 9 RBLs but 3 XBLs.
 We think that
whether or not the radio polarization in XBLs is higher than that in
RBLs, more radio polarization observations are significant and
encouraged.

\begin{figure}
\begin{center}
    \FigureFile(90mm,100mm){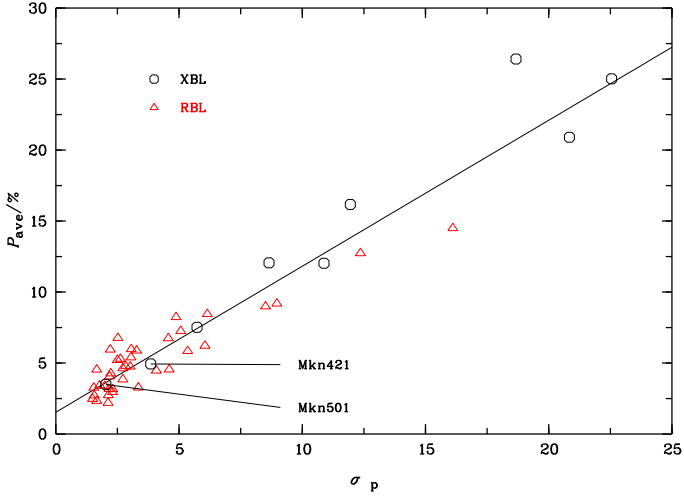}
\end{center}
\caption{ The relation between the
averaged polarization at 8 GHz and the polarization
variability--$\sigma_{\rm{P}}$ for radio selected BL Lacertae
objects--RBL (triangles) and X-ray selected BL Lacertae objects--XBL
(open circles). The line stands for the best fitting result.}
\label{Fan-2006-PASJ-fg-PDP}
\end{figure}

When we take into account of the polarization variation, we can use
the standard deviation, $\sigma_{\rm{P}}$ as the variation
indication. In this sense, we found that the variation  is closely
linearly correlated with the averaged polarization shown in Fig.
\ref{Fan-2006-PASJ-fg-PDP}, the line stands for the best fitting
result,
$P_{\rm{ave}}(\%)=(1.02\pm0.04)\sigma_{\rm{P}}+(1.54\pm0.32)$. Here
we used the data at 8GHz.

When we considered the relationship of polarization on the spectral
index, we found that the radio polarization increases with the
spectral
   index, $\alpha$ ($S_{\nu} \propto \nu^{-\alpha}$). The
 data can
gave a linear relation of $P \sim 9.15\alpha + 7.04$. From the
synchrotron emission, one can get that the polarization is
proportional to $\frac{\alpha+1}{\alpha+5/3}$, which suggests that
the polarization is correlated with the spectral index and that the
polarization increases with the spectral index.
   Although our statistical result show a trend that the polarization
increase with the radio spectral index as shown in Fig. 1, the trend
is largely different from the prediction by synchrotron emission.

Polarization  is found to be associated with the core-dominance
parameter ( see Wills et al. 1992 and reference therein) with high
polarization corresponding to large $\log R$.  For the present
sample,  one can get a plot of the radio polarization against the
core-dominance parameter as shown in Fig. \ref{Fan-2006-PASJ-fg-PR}.
In Fig. \ref{Fan-2006-PASJ-fg-PR}, we only showed the averaged radio
polarization at 8.0GHz and the core dominance parameters that are
available in the literature.

Based on the two-component beaming model Urry \& Shafer (1984) (see
also Urry \& Padovani, 1995), the emissions of an AGN are  from  two
components, namely the beamed and the unbeamed ones.
 Assuming  the intrinsic flux of the jet,
 $S_{j}^{in}$, to be a fraction $f$ of the unbeamed flux,
 $S_{unb}$, i.e.,
 $S_{j}^{in} = fS_{unb}$, one can get
 $ S^{ob} = S_{unb} + S_{j}^{ob} = (1 + f\delta^{p}) S_{unb}$.
 If one assumes that the emissions in the co-moving jet are also
 composed of two components, namely the polarized and the unpolarized, with
 the two components being proportional to each other, $S_{j}^{in} = S_{j}^{p} +
 S_{j}^{up}$, $S_{j}^{p} = \eta S_{j}^{up}$. Then  a relation
 between the observed polarization and the Doppler factor can be obtained
(Fan et al. 1997; see also Fan et al. 2001).

\begin{equation}
P^{ob} = {\frac{(1 + f) \delta^{p}}{1 + f\delta^{p}}} P^{in}
\end{equation}
where $P^{in}$ is the intrinsic polarization expressed in the form
\begin{equation}
P^{in} = {\frac{f}{1 + f}}{\frac{\eta}{1 + \eta}}
\end{equation}
which follow
 \begin{equation}
P^{ob} = {\frac{f \delta^{p}}{1 + f\delta^{p}}}
{\frac{\eta}{1+\eta}}
 \end{equation}
 $\delta$ is the  Doppler factor,  $\eta$ is   the ratio of the polarized to
 the unpolarized luminosity in the jets. The value
 of $p$ depends on the shape  of the emitted spectrum and the detailed
 physics of the jet (Lind \& Blandford 1985), $p = 3 + \alpha$ is for
 a moving sphere and $ p = 2 + \alpha$  is for the case of a continuous
 jet,  $\alpha$ is the spectral index.

 If we use the expression for
 the  core-dominance parameter,  $R=f((\Gamma(1-\beta cos \theta))^{-p}+(\Gamma(1+\beta cos
\theta))^{-p})$=$f\delta^p[1+f(\frac{1-\beta cos \theta}{1+\beta cos
\theta})^p]$, then the polarization can be written in the form

\begin{equation}
P^{ob} = {\frac{R-G(f,\theta,\beta)}{1-G(f,\theta,\beta)+R}}
{\frac{\eta}{1+\eta}},
 \end{equation}
here, $G(f,\theta,\beta)=f(\frac{1-\beta cos \theta}{1+\beta cos
\theta})^p$. Since the observed polarization is positive, then we
can get the value of $G(f,\theta,\beta)$ to be smaller than the
value of R. Therefore, the value of $G(f,\theta,\beta)$ is actually
very small since the $\log R$ is -0.91 for 1100+772. So, we have
\begin{equation}
P^{ob} \sim {\frac{R}{1+R}} {\frac{\eta}{1+\eta}}.
 \end{equation}
For a given $\eta$, a theoretical relation of polarization depending
on the core-dominance parameter can be obtained. In Fig.
\ref{Fan-2006-PASJ-fg-PR}, we showed the polarization vs
core-dominance parameter
   plot.
   The core-dominance  parameter is taken as the jet viewing indication, it is actually the
 indication of a beaming effect. For a strongly boosted source, the core-dominance parameter can
 be simply expressed as $R=f\delta^p$ (Ghisellini et al. 1993).
 In this sense, one can  see clearly that 1) the polarization is consistent with the
explanation that the polarization is associated with the beaming
effect and 2) for most sources, their $\eta$ is less than 0.25.

In this paper, based on the MURAO data base, we considered the radio
polarization for BL Lacertae objects and found that the averaged
polarization depends on the frequency, the spectral index and the
core-dominance parameter. The polarization variation is closely
correlated with the polarization.
    The relativistic beaming could explain the
polarization characteristic of BL Lacs.

\section{Acknowledgements}
  We thank the referee for the useful comments and suggestions.
This work is partially supported by the National 973 project (NKBRSF
G19990754), the National Science Fund for Distinguished Young
Scholars (10125313), the National Natural Science Foundation of
China (10573005,10633010), and the Fund for Top Scholars of
Guangdong Province (Q02114). We also thank the financial support
from the Guangzhou Education Bureau and Guangzhou Science and
Technology Bureau. This research has made use of data from the
University of Michigan Radio Astronomy Observatory which has been
supported by the University of Michigan and the National Science
Foundation.

\begin{table*}
 \begin{center}
 Table~ 1. \hspace{12pt} A sample of 47 BL Lacertae objects
  \vspace{12pt}
\begin{tabular}{lccccccc}
\hline\\
Source & Class& $\alpha$ & $P_{8GHz}^{ave.}$ & $\sigma_p$& Data span
&$\log R$ & Ref
\\\hline

0003-066    &   RBL    &   0.088   &   3.27    &   1.54    &1978-1998  &         &       \\
0048-097    &   RBL    &   -0.242  &   5.4     &   3.05    &1970-1999  &1.3      &   G93 \\
0109+224    &   XBL    &   -0.188  &   7.52    &   5.73    &1978-1999  &1.62     &   W92 \\
0215+015    &   RBL    &   -0.337  &   3.16    &   2.31    &1979-1999  &0.9      &   W92 \\
0219+428    &   RBL    &   0.754   &   2.96    &   2.3     &1974-1999  &0.3      &   G93 \\
0235+164    &   RBL    &   -0.192  &   2.74    &   2.13    &1974-1999  &2.2      &   G93 \\
0300+470    &   RBL    &   -0.058  &   3.38    &   1.99    &1975-1999  & 1.9     &   G93 \\
0323+022    &   XBL    &   0.252   &   25.02   &   22.55   &1985-1999  &         &       \\
0422+004    &   RBL    &   -0.191  &   5.85    &   5.35    &1978-1999  & 3.3     &   W92 \\
0716+714    &   RBL    &   -0.271  &   9.19    &   8.97    &1981-1999  & -0.52   &   W92 \\
0735+178    &   RBL    &   0.062   &   2.48    &   1.49    &1977-1999  & 3.4     &   G93 \\
0754+100    &   RBL    &   -0.09   &   5.3     &   2.62    &1978-1999  & 1.14    &   W92 \\
0808+019    &   RBL    &   -0.133  &   5.98    &   3.06    &1979-1999  & 1.55    &   W92 \\
0814+425    &   RBL    &   -0.035  &   4.06    &   2.18    &1977-1999  & 0.9     &   W92 \\
0818-128    &   RBL    &   -0.054  &   4.47    &   4.08    &1979-1999  & 0.23    &   W92 \\
0829+046    &   RBL    &   -0.239  &   3.42    &   1.78    &1978-1999  & 0.71    &   F03 \\
0851+202    &   RBL    &   -0.385  &   4.78    &   2.83    &1971-1999  & 3.5     &   G93 \\
0912+297    &   XBL    &   0.318   &   12.06   &   8.65    &1980-1999  & 0.3     &   W92 \\
0954+658    &   RBL    &   -0.022  &   14.5    &   16.11   &1974-1999  & 0.53    &   W92 \\
0957+227    &   RBL    &   0.757   &   12.74   &   12.36   &1979-1999  &         &       \\
1100+772    &   XBL    &   0.625   &   16.17   &   11.95   &1983-1999  & -0.91   &   W92 \\
1101+384    &   Mid-BL &   0.182   &   4.94    &   3.85    &1978-1999  & 0.76    &   F03 \\
1133+704    &   XBL    &   0.184   &   26.41   &   18.67   &1980-1999  &         &       \\
1147+245    &   RBL    &   0.023   &   3.85    &   2.72    &1979-1999  & 1.42    &   W92 \\
1215+303    &   RBL    &   -0.008  &   6.22    &   6.05    &1979-1999  & 0.27    &   W92 \\
1219+285    &   RBL    &   -0.065  &   4.75    &   3.02    &1978-1999  & -0.72   &   F03 \\
1307+121    &   RBL    &   -0.002  &   5.88    &   3.28    &1978-1999  &         &       \\
1308+326    &   RBL    &   -0.205  &   2.72    &   1.56    &1976-1999  & 1.7     &   G93 \\
1400+162    &   RBL    &   0.481   &   8.99    &   8.51    &1978-1999  & -0.11   &   F03 \\
1413+135    &   RBL    &   -0.601  &   2.2     &   2.12    &1978-1999  & 0.55    &   F03 \\
1418+546    &   RBL    &   -0.117  &   3.28    &   3.35    &1978-1999  & 1.77    &   W92 \\
1514+197    &   RBL    &   -0.065  &   6.75    &   4.57    &1978-1999  & 2.18    &   W92 \\
1538+149    &   RBL    &   -0.147  &   5.95    &   2.21    &1977-1999  & 0.95    &   W92 \\
1652+398    &   Mid-BL &   0.208   &   3.5     &   2.03    &1977-1999  & 0.07    &   F03 \\
1717+178    &   RBL    &   -0.127  &   7.26    &   5.07    &1977-1999  & 1.78    &   W92 \\
1727+502    &   XBL    &   0.41    &   20.9    &   20.84   &1979-1999  & 1.01    &   W92 \\
1749+096    &   RBL    &   -0.473  &   3.18    &   2.15    &1978-1999  & 0.5     &   F03 \\
1749+701    &   RBL    &   0.051   &   8.45    &   6.15    &1980-1999  & -0.46   &   F03 \\
1803+784    &   RBL    &   -0.102  &   4.27    &   2.23    &1981-1999  & 0.42    &   F03 \\
1807+698    &   RBL    &   0.048   &   4.54    &   1.66    &1979-1999  & 0.68    &   F03 \\
1823+568    &   RBL    &   -0.099  &   6.77    &   2.52    &1981-1999  &         &       \\
2007+777    &   RBL    &   -0.091  &   5.22    &   2.5     &1981-1999  & 1.36    &   G93 \\
2032+107    &   RBL    &   0.116   &   4.55    &   4.6     &1978-1999  & 1.61    &   W92 \\
2131-021    &   RBL    &   0.065   &   2.34    &   1.66    &1974-1999  &         &       \\
2155-304    &   XBL    &   0.097   &   12.03   &   10.88   &1979-1999  & 0.66    &   W92 \\
2200+420    &   RBL    &   -0.019  &   4.63    &   2.72    &1968-1999  & -0.05   &   F03 \\
2254+074    &   RBL    &   0       &   8.24    &   4.88    &1979-1999  & 1.77    &   W92 \\

\hline
\end{tabular}
\\
Note to the Table:\
  Col. 1 gives the name of the source,
  Col. 2 the classification, XBL is for X-ray selected BL Lacertae
  object, RBL for radio selected BL Lacertae object while mid-term
  for the object between XBL and RBL,
  Col. 3 for the spectral index($S_{\nu}\,\propto\nu^{-\alpha}$),
  Col. 4 for the averaged observation polarization at 8GHz,
  Col. 5 for the 1 $\sigma$ deviation of the averaged polarization
  in Col. 4,
  Col. 6 for the time span of the observation data,
  Col. 7 for the core-dominance parameter,
  Col. 8 reference for the core-dominance parameter in Col. 7,
  respectively.\\
 F03: Fan \& Zhang (2003);
 G93: Ghisellini et al. (1993);
 W92: Wills et al. (1992)
\end{center}
\end{table*}

\end{document}